# Investigation of Hopping Conduction and Its Impact on the Subthreshold and Transition Region Transport in Oxide Semiconductor Transistors by Magneto-Transport Measurements

Chen Wang[1], Liankai Zheng[1], Kai Jiang[1], Jinxiu Zhao[1], Zhenyu Zhang[2], Xuefei Li[2], and Mengwei Si[1,*]

[1] National Key Laboratory of Advanced Micro and Nano Manufacture Technology and School of Information Science and Electronic Engineering, Shanghai Jiao Tong University, Shanghai 200240, China

[2] School of Integrated Circuits, Huazhong University of Science and Technology, Wuhan 430074, China

* Corresponding author. Email: mengwei.si@sjtu.edu.cn

## Abstract

In this work, magneto-transport measurements, including Hall effects and magnetoresistance (MR) at various temperatures, are employed to directly probe the electron transport properties in crystalline indium oxide ($In_2O_3$) and amorphous indium-zinc oxide (IZO) transistors. For the first time, we develop a subgap density of states (DOS) extraction method based on the MR measurements at low temperature, considering both the interference and orbital shrinkage effects. The sign and magnitude of MR are used as direct evidence to distinguish the dominating transport mechanisms between hopping conduction and free-electron conduction. It is found that crystalline $In_2O_3$ exhibits a subgap DOS more than two orders of magnitude lower than that of amorphous IZO, together with a smaller localization radius of hopping sites. As a result, crystalline $In_2O_3$ not only enhances the carrier mobility but also significantly reduces the supply voltage ($V_{DD}$) because of the smaller transition region enabled by the much-suppressed subgap DOS. This study reveals that structural disorder critically influences the device operation in the subthreshold and transition regions of oxide semiconductor transistors.

**KEYWORDS:** oxide semiconductor, hopping conduction, magnetoresistance, density of states, subthreshold

Oxide semiconductors (OS) have emerged as promising channel materials for monolithic three-dimensional (3D) integration and dynamic random-access memory (DRAM) applications[1–7]. As the operating voltage of OS transistors is scaled down toward the $V_{DD}$ requirements of advanced nodes, the device operation in the subthreshold and transition regions becomes increasingly important. However, in contrast to Si MOSFETs[8–12], the fundamental carrier transport mechanisms governing the subthreshold and transition regions in OS transistors remain unclear. Several fundamental challenges persist in modeling and characterizing OS transistors in these operational regimes. First, distinguishing the dominating transport mechanisms, such as free-electron conduction, percolation conduction[13], and hopping conduction[14], across different gate voltage ($V_{GS}$) ranges remains ambiguous. Second, accurate modeling of the subgap density of states (DOS) is challenging due to the absence of direct experimental probes[15]. Third, the impact of the material structure of OS, such as amorphous and crystalline phases, on the subgap DOS and the transport behavior in these regions has not been clarified.

To address these challenges, we employ magneto-transport measurements, including gated-Hall and magnetoresistance (MR) measurements, to directly probe the electron transport in OS transistors[16–18]. The gated-Hall measurement avoids the influence of contact resistance and enables the simultaneous acquisition of the Hall mobility ($\mu_{\mathrm{Hall}}$), the Hall carrier density ($n_{\mathrm{Hall}}$), and the MR of the same device. Based on the variable-range-hopping (VRH) MR model, we clarify that hopping conduction dominates the electron transport in the transition and subthreshold regions of OS transistors[19,20]. From the MR fitting, the subgap DOS and the localization radius of hopping sites are directly extracted. Furthermore, we reveal that the subgap DOS of crystalline $In_2O_3$ is significantly lower than that of amorphous IZO, which arises from the difference in structural disorder. As a result, the crystalline $In_2O_3$ device exhibits a much smaller transition region, which is favorable for low-$V_{DD}$ operation.

Figs. 1(a) and 1(b) show the schematic diagram and the fabrication process flow of the bottom-gated $In_2O_3$/IZO transistors used in this work. The OS transistors consist of 40-nm Mo as the gate metal, 8-nm $HfO_2$ grown by atomic layer deposition (ALD) as the gate dielectric, 3-nm $In_2O_3$ or 8-nm IZO grown by ALD as the semiconducting channel, 40-nm ITO deposited by sputtering as the source/drain (S/D) electrodes, and 8-nm $HfO_2$ grown by ALD as the passivation layer. The fabricated devices were annealed in pure $O_2$ at 400 °C for 20 min. The fabrication process is similar to that reported in ref [16]. Fig. 1(c) shows the cross-sectional scanning transmission electron microscopy (STEM) image with energy-dispersive X-ray spectroscopy (EDS) mapping at the channel region of an $In_2O_3$ transistor, clearly capturing the $HfO_2$/$In_2O_3$/$HfO_2$ stack. Fig. 1(d) presents the grazing-incidence X-ray diffraction (GIXRD) patterns of the as-deposited and annealed $HfO_2$/$In_2O_3$/$HfO_2$ and $HfO_2$/IZO/$HfO_2$ stacks. The $In_2O_3$ channel undergoes an obvious crystallization after annealing, whereas the IZO channel remains amorphous, providing a pair of model systems with distinct structural disorder for the transport study.

Figs. 1(e) and 1(f) show the transfer characteristics ($I_D$–$V_{GS}$) of the $In_2O_3$ and IZO transistors at $V_{DS}$ of 1 V measured from 4 K to 300 K. The field-effect mobility ($\mu_{\mathrm{FE}}$) extracted at $V_{DS}$ of 0.1 V is plotted in Figs. 1(g) and 1(h) for the $In_2O_3$ and IZO transistors as a function of temperature. The $\mu_{\mathrm{FE}}$ of the $In_2O_3$ transistor reaches 145.6 $cm^2/V{\cdot}s$ at 300 K and further improves to 160.7 $cm^2/V{\cdot}s$ at 4 K, indicating the significantly enhanced carrier transport in the crystalline channel. It has been previously demonstrated that the deposition of the top $HfO_2$ dielectric above the $In_2O_3$ channel significantly enhances its crystallization, exhibiting an epitaxy-like growth behavior that facilitates efficient carrier transport[16]. In contrast, the $\mu_{\mathrm{FE}}$ of the IZO transistor is 42.4 $cm^2/V \cdot s$ at 300 K and 39.4 $cm^2/V \cdot s$ at 4 K, which is limited by the amorphous structure.

Fig. 2 analyzes the subthreshold behavior of the devices. Figs. 2(a) and 2(b) plot the subthreshold swing (SS) versus $I_D$ of the $In_2O_3$ and IZO transistors measured at 4, 100, 200, and 300 K with $L_{ch}$ from 2 to 80 μm at $V_{DS}$ of 1 V. Transistors with different $L_{ch}$ exhibit near-identical SS at the same temperature, indicating that the subthreshold characteristics are governed by the channel and the gate stack rather than by channel-length-dependent effects. The temperature-dependent SS of the $In_2O_3$ and IZO transistors with $L_{ch}$ of 2 μm extracted from 4 K to 300 K at $V_{DS}$ of 1 V is shown in Figs. 2(c) and 2(d). The SS of both devices saturates below 100 K, indicating the presence of exponential band-tail states[8]. Such band-tail-limited saturation of SS at cryogenic temperatures has been widely observed and theoretically analyzed, and recent studies show that SS below 1 K is governed by the detailed shape of the band-tail DOS, with hybrid band tails containing both traps and mobile states determining the low-temperature SS behavior[11].

The gated-Hall bar devices were fabricated together with the transistors using the same process flow, where the length and width of the Hall bar are 360 μm and 60 μm. For the gated Hall effect measurement, $\mu_{\mathrm{Hall}}$ and $n_{\mathrm{Hall}}$ are acquired as functions of $V_{GS}$. The Hall bar geometry satisfies the requirements of the ASTM F76 standard, and the measurement setup is schematically illustrated and described in Fig. S1. Figs. 3(a) and 3(b) present the voltage drop parallel to the channel ($V_{XX}$) and the Hall voltage ($V_{XY}$) versus $V_{GS}$ characteristics of an $In_2O_3$ Hall bar device measured at 2 K with magnetic field ($B$) from −14 T to 14 T. The extraction of $n_{\mathrm{Hall}}$ and $\mu_{\mathrm{Hall}}$ is meaningful only while the Hall bar is turned on. In the off state, $V_{XX}$ and $V_{XY}$ no longer respond to the magnetic field and no reliable Hall signal can be measured, so that the extracted carrier density is not physical. The critical $V_{GS}$ is therefore identified as the value at which $V_{XX}$ starts to decrease, and the extraction is restricted to $V_{GS}$ values above it. At 2 K this critical $V_{GS}$ is −0.8 V for the $In_2O_3$ device (Fig. 3(a)) and −1.2 V for the IZO device (Fig. S2). All $n_{\mathrm{Hall}}$ and $\mu_{\mathrm{Hall}}$ data reported here are taken above these values, which ensures the accuracy of the Hall measurement. Figs. 3(c) and 3(d) plot the extracted $\mu_{\mathrm{Hall}}$ and $n_{\mathrm{Hall}}$ of the $In_2O_3$ device at 2, 100, 200, and 300 K. The $\mu_{\mathrm{Hall}}$ of the $In_2O_3$ device reaches 105.1 $cm^2/V\cdot s$ at 300 K and 155.1 $cm^2/V\cdot s$ at 2 K. The $n_{\mathrm{Hall}}$ of the $In_2O_3$ device increases linearly with $V_{GS}$ and follows $n_{\mathrm{Hall}} = \mathrm{C_{ox}}(\mathrm{V_{GS}} - \mathrm{V_{TH}})/\mathrm{q}$, where $\mathrm{C_{ox}}$ is the gate capacitance per unit area and $V_{TH}$ the threshold voltage, $q$ is the elementary charge.

Figs. 3(e) and 3(f) plot the extracted $\mu_{\mathrm{Hall}}$ and $n_{\mathrm{Hall}}$ of the IZO device at the same temperatures. The $\mu_{\mathrm{Hall}}$ of the IZO device reaches 39.1 $cm^2/V\cdot s$ at 300 K and 41.8 $cm^2/V\cdot s$ at 2 K, which is limited by the amorphous structure. In contrast to the linear $n_{\mathrm{Hall}}$–$V_{GS}$ of the crystalline $In_2O_3$ device, the $n_{\mathrm{Hall}}$ of the IZO device exhibits an obvious non-linearity at 2 K, which indicates that the measured $n_{\mathrm{Hall}}$ is governed by the mobility of two distinct carriers, rather than by a linear superposition of carrier densities, according to the two-carrier Hall effect model[21]. In a system containing both free electrons and localized (hopping) carriers with different mobilities, the measured $n_{\mathrm{Hall}}$ is a mobility-weighted combination of the two carrier species rather than their sum. In the low-mobility limit ($\mu B \ll 1$), it can be expressed as

$$n_{\mathrm{Hall}} = \frac{(n_{\mathrm{Free}}\mu_{\mathrm{Free}}+n_{\mathrm{Loc}}\mu_{\mathrm{Loc}})^2}{n_{\mathrm{Free}}\mu_{\mathrm{Free}}^2+n_{\mathrm{Loc}}\mu_{\mathrm{Loc}}^2} \tag{1}$$

where $n_{\mathrm{Free}}$ and $n_{\mathrm{Loc}}$ are the densities of the free and localized carriers, $\mu_{\mathrm{Free}}$ and $\mu_{\mathrm{Loc}}$ are their respective mobilities. Therefore, when hopping conduction coexists with free-electron conduction, $n_{\mathrm{Hall}}$ deviates from $n_{\mathrm{Free}}+n_{\mathrm{Loc}}$, and the deviation becomes more pronounced at low temperature where the two mobilities differ greatly. As demonstrated by two-carrier Hall studies in other material systems[21], a single-carrier interpretation of such measurements always underestimates the total carrier density and overestimates the carrier mobility. The error magnifies with increasing mobility ratio $\mu_{\mathrm{Loc}}/\mu_{\mathrm{Free}}$ and is most pronounced at low temperatures, consistent with the strong non-linearity observed at 2 K.

The SS of the IZO transistor saturates at low temperature, indicating an exponential energy distribution of the localized states. This alone cannot reproduce the measured $n_{\mathrm{Hall}}$–$V_{GS}$ during numerical fitting process, and the subgap DOS was therefore described by an exponential band tail combined with a Gaussian distribution (Fig. S3). On the basis of this DOS, $n_{\mathrm{Free}}$ and $n_{\mathrm{Loc}}$ were calculated as functions of $V_{GS}$ (section 3 in Supporting Information) and substituted into eqn. (1). The resulting numerical $n_{\mathrm{Hall}}$–$V_{GS}$ reproduces the experimental data of the IZO device at 2 K, indicating a high density of localized states in the amorphous channel. It should be noted, however, that this DOS is inferred from $n_{Hall}$ measured in the on state so it thus provides only an approximate description of DOS.

To directly probe the hopping conduction, the transfer curves of the $In_2O_3$ and IZO Hall bar devices measured at 2 K and $B$ of 0 T are plotted in Figs. 4(a) and 4(e), where MR is defined as $MR = [R(B) - R(0)]/R(0)$. For the MR measurements, two Keithley 2450 high-precision source-measure units (SMUs) were used to apply $V_{GS}$ and $V_{SD}$ and to measure $I_G$ and $I_D$, respectively, while the $V_{XX}$ and $V_{XY}$ sensing terminals were left disconnected, so that the small subthreshold currents could be measured accurately without the loading effect of the lock-in amplifier input impedance (Fig. S4). Based on the behavior of the MR, each transfer curve is divided into three regions, and the MR measured with $B$ from −14 T to 14 T within each region is shown in Figs. 4(b)–4(d) for the $In_2O_3$ Hall bar (regions I, II, and III, respectively) and in Figs. 4(f)–4(h) for the IZO Hall bar. At low $B$, a negative MR occurs, whereas at high $B$, a positive MR can be observed. In regions I and II, the positive MR appears at high $B$ and is suppressed with increasing $V_{GS}$, whereas only a negative MR occurs in region III. The positive MR of the IZO device is much more pronounced in regions I and II than that of the $In_2O_3$ device.

The MR in region III at high $V_{GS}$ can be well explained by the weak localization of free electrons, suggesting that free-electron transport dominates at high $V_{GS}$, as expected[22]. In contrast, the MR in regions I and II can be well described by the VRH-induced MR model based on the single-scattering-path approach[20], suggesting that hopping transport dominates at low $V_{GS}$. In this approach, the VRH-induced MR consists of two competing contributions, negative MR and positive MR. In the single-scattering-path model, the interfering tunneling paths of a hopping site are confined within a cigar-shaped region of length $r$ (the hopping length) and width $(ar)^{1/2}$. In a magnetic field, each path acquires a field-dependent phase, so that the destructive interference among the paths is suppressed and the tunneling probability is enhanced, giving rise to a negative MR that grows linearly with $B$ at low fields. In contrast, the positive MR at high magnetic fields originates from the orbital shrinkage of the localized wave functions, which reduces the overlap between adjacent hopping sites and suppresses the tunneling probability. In the weak-field regime, the shrinkage makes the inter-site overlap decrease quadratically with $B$, so that the positive MR increases as $B^2$. The total MR is obtained by summing the two contributions[20]:

$$\frac{\Delta R(B)}{R(0)} = \frac{\Delta R(B)}{R(0)}|_{\mathrm{Int}} + \frac{\Delta R(B)}{R(0)}|_{\mathrm{Orb}} \tag{2}$$

with

$$\frac{\Delta R(B)}{R(0)} = -\frac{2}{\pi}\int_0^{\pi/2} d\varphi \ln[\frac{\sin^2\varphi + G^2(\varphi,B)}{\sin^2\varphi + G^2(\varphi,0)}] - 1 \tag{3}$$

$$G(\varphi,B) = \pi A(\frac{B}{B_0})F(\frac{\varphi B_0}{B}) \tag{4}$$

$$F(x) = \frac{2}{\pi^{3/2}}\int_{-\infty}^{\infty} dt e^{-t^2}\int_{-\pi/2}^{\pi/2} d\alpha(\cos^2\alpha)|x - t\cos\alpha| \tag{5}$$

$$\frac{\Delta R(B)}{R(0)} = \frac{5}{2016}\frac{q^2 a^4}{\hbar^2}B^2(\frac{T_0}{T})^{3/4} \tag{6}$$

where $a$ the localization radius of the hopping sites, DOS the density of states at the Fermi level, and $V_0$ the amplitude prefactor of the inter-site overlap integral. The derived quantities are the Mott temperature $T_0 = 14/(DOS \cdot a^2)$, the maximum hopping length $r_c = (a/2)(T_0/T)^{1/3}$, the scattering intensity $A = (\pi^{3/2}/2^{5/2})DOS \cdot V_0\sqrt{ar_c^3}$, and the characteristic field $B_0 = (\sqrt{2}\varphi_0)/(\pi\sqrt{ar_c^3})$, with $\varphi_0$ the flux quantum.

By fitting the experimental MR with the localization radius $a$, the DOS, and the overlap-integral prefactor $V_0$ as fitting variables, the DOS and $a$ versus $V_{GS}$ can be acquired (Fig. S5), as shown in Fig. 5(a) for the $In_2O_3$ device and Fig. 5(b) for the IZO device. Unlike the two-carrier analysis of $n_{\mathrm{Hall}}$, which relies on on-state data and on assumptions about the carrier mobilities, the MR measurements are carried out in the subthreshold region and thus probe the deep defect states directly. The VRH MR model contains no assumption regarding the carrier mobility, since the DOS and the localization radius enter it directly; the subgap DOS obtained from the MR analysis is therefore expected to be more accurate than that inferred from the two-carrier fit of $n_{Hall}$. What the MR analysis yields directly is the subgap DOS as a function of $V_{GS}$, which is then converted onto an energy scale by mapping $V_{GS}$ onto the surface potential $\psi_s$ (section 5 in Supporting Information). The Fermi level $E_F = q\psi_s$ then gives the energy position of the states being probed, so that the DOS and the localization radius $a$ are obtained as functions of the energy $E - E_C$, as shown in Figs. 5(c) and 5(d). The extracted subgap DOS can be decomposed into two distinct components: an exponential tail component and a Gaussian tail component. The exponential tail is primarily responsible for the saturation of the SS at low temperature. The total DOS in the IZO device is more than two orders of magnitude larger than that in the $In_2O_3$ device (Figs. 5(c) and 5(d)), which most likely results from the amorphous structure of IZO. In addition, the larger localization radius $a$ of the IZO device (Fig. 5(e)) indicates that the wave functions of the hopping states are more extended in the amorphous material[23].

The transition region between the subthreshold and above-threshold operation has been identified as a critical metric for quantifying the supply-voltage scalability of OS transistors[24]. Fig. 5(f) compares the transition-region width $V_{TR} = V_{ON} - V_{OFF}$ of the crystalline $In_2O_3$ and amorphous IZO transistors, where $V_{OFF}$ and $V_{ON}$ are extracted when $I_D$ is equal to $10^{-10}$ A/μm and $10^{-6}$ A/μm, respectively. $V_{TR}$ is 0.91 V for the amorphous IZO transistor, significantly wider than 0.61 V for the crystalline $In_2O_3$ transistor. This wider $V_{TR}$ directly reflects a broader transition region in the IZO device, which originates from its significantly higher subgap DOS (Figs. 5(c) and 5(d)) acting as the shallow trap states that broaden the transition region. Consequently, employing a crystalline channel such as $In_2O_3$ not only enhances the carrier mobility but also reduces $V_{DD}$, enabled by the much-suppressed localized states below the conduction-band edge.

In summary, for the first time, this work directly measures the hopping conduction and the corresponding subgap DOS and localization radius of hopping sites in OS transistors using magneto-transport measurements. We reveal the critical role of structural disorder on the electron transport in the subthreshold and transition regions of OS transistors. It is understood that employing a crystalline channel such as $In_2O_3$, compared with amorphous OS, not only enhances the carrier mobility but also can reduce $V_{DD}$ because of the smaller transition region enabled by the much-suppressed localized states below $E_C$. These findings provide direct experimental insight into the fundamental transport physics of OS transistors and offer guidance for the design of low-voltage oxide electronics.

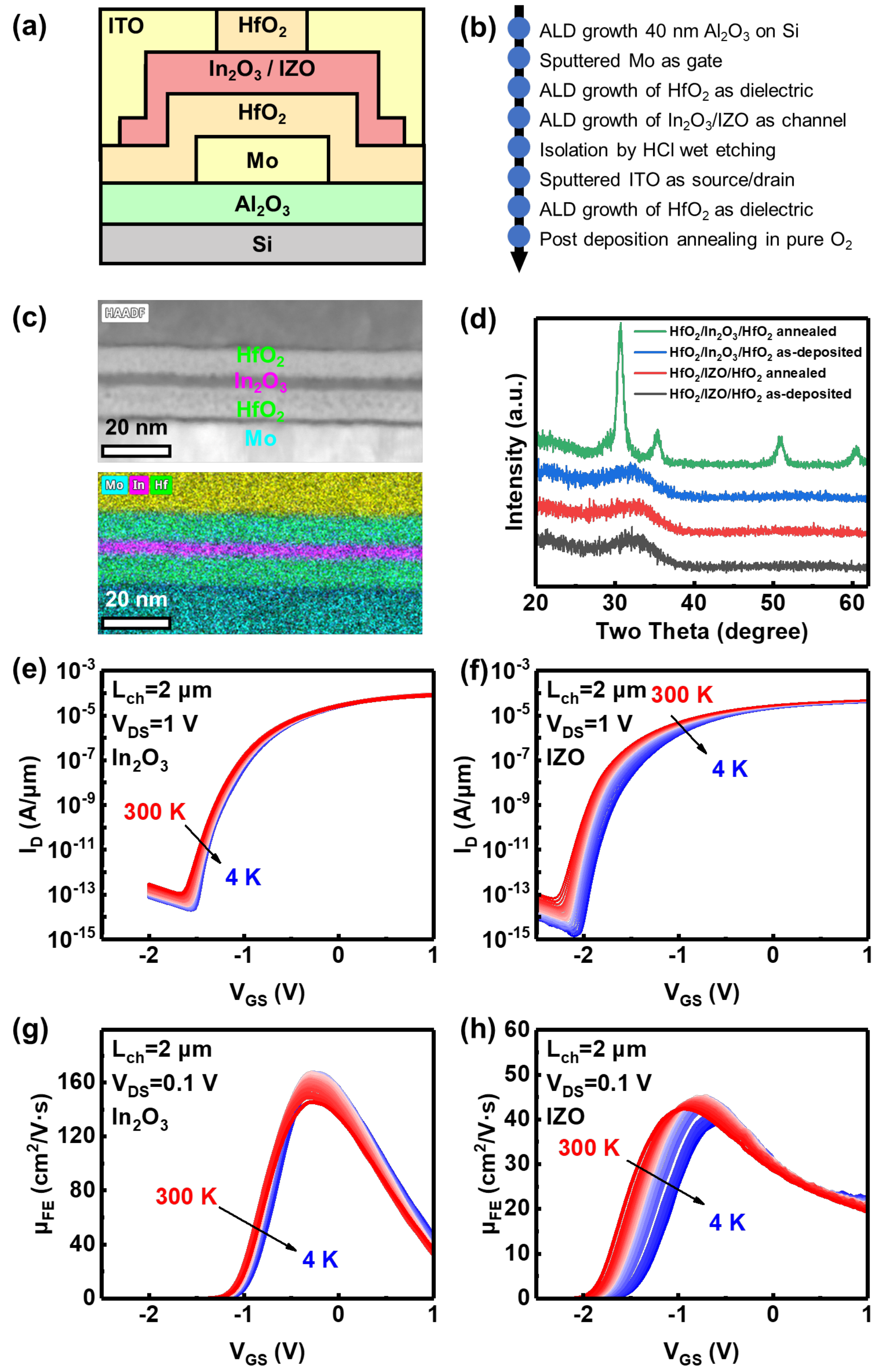

**Figure 1.** (a) Schematic diagram and (b) fabrication process flow of bottom-gated $In_2O_3$/IZO transistors. (c) Cross-sectional STEM image with EDS mapping of the $In_2O_3$

transistor, showing a clear $HfO_2/In_2O_3/HfO_2$ stack. (d) GIXRD patterns of as-deposited/annealed $HfO_2/In_2O_3/HfO_2$ and $HfO_2$/IZO/$HfO_2$ stacks. The $In_2O_3$ shows an obvious crystallization after annealing but the IZO remains amorphous. Transfer curves of (e) $In_2O_3$ and (f) IZO transistors at $V_{DS}$ of 1 V from 4 K to 300 K. Extracted $\mu_{FE}$ of (g) $In_2O_3$ and (h) IZO transistors at $V_{DS}$ of 0.1 V from 4 K to 300 K.

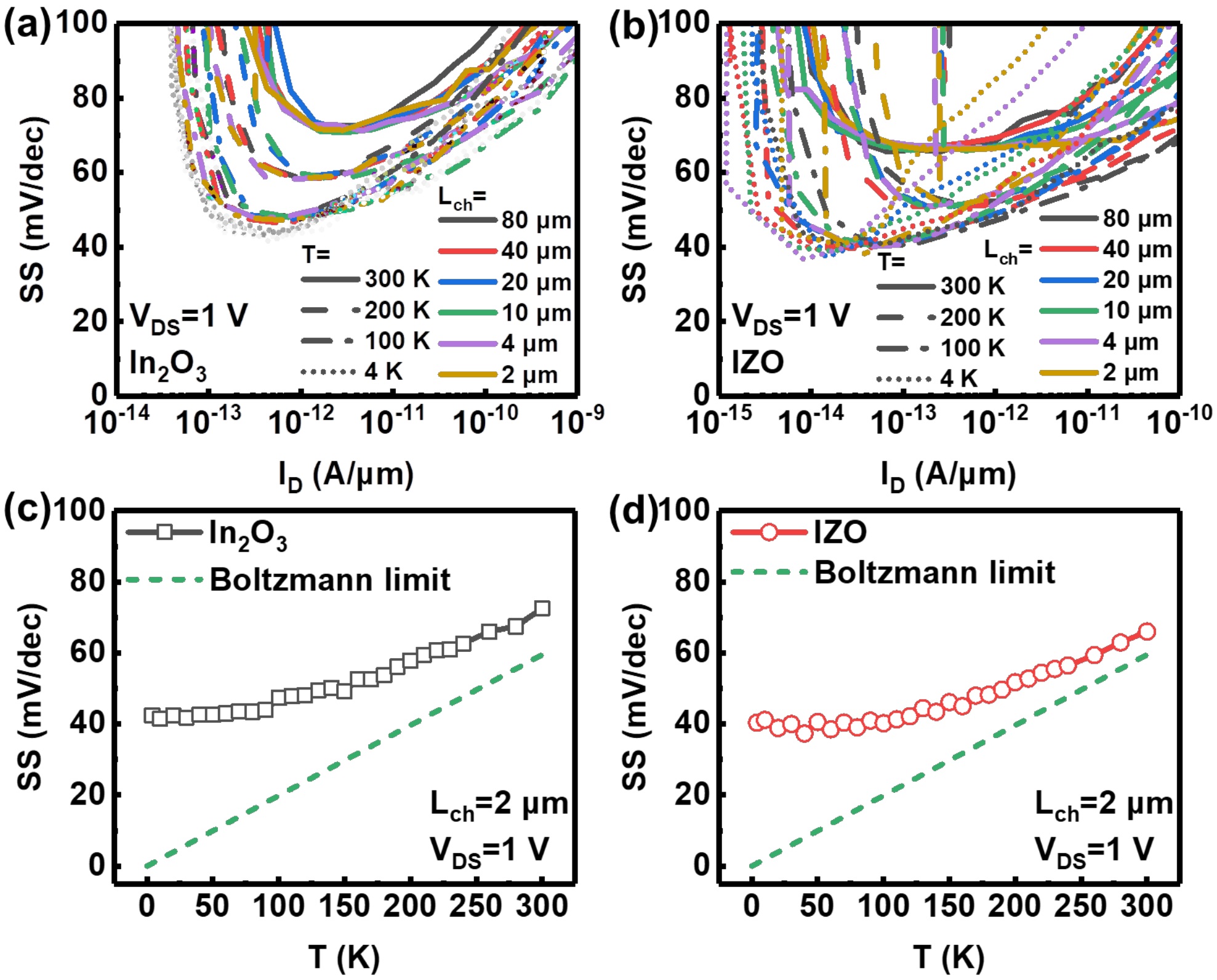

**Figure 2.** The SS versus $I_D$ plots of (a) $In_2O_3$ and (b) IZO under 4, 100, 200, and 300 K with $L_{ch}$ of 2–80 μm at $V_{DS}$ of 1 V. Transistors with different $L_{ch}$ have near-identical SS at the same temperature. Extracted SS from 4 K to 300 K of (c) $In_2O_3$ and (d) IZO transistors with $L_{ch}$ of 2 μm at $V_{DS}$ of 1 V.

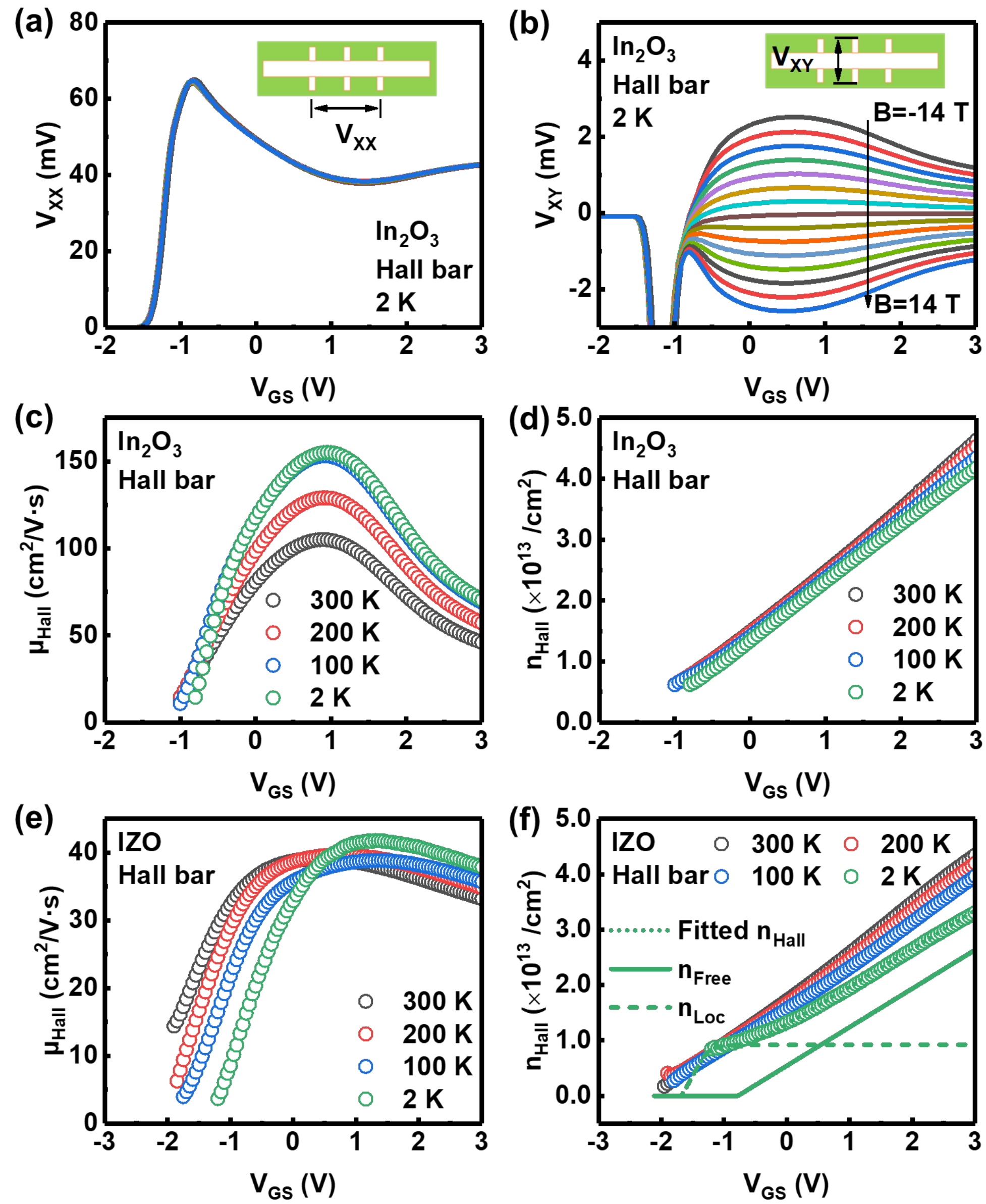

**Figure 3.** (a) $V_{XX}$ and (b) $V_{XY}$ versus $V_{GS}$ characteristics of an $In_2O_3$ Hall bar measured at 2 K with B from −14 T to 14 T. Extracted $\mu_{Hall}$ of (c) the $In_2O_3$ and (e) the IZO Hall bar, and extracted $n_{Hall}$ of (d) the $In_2O_3$ and (f) the IZO Hall bar, at 2, 100, 200, and 300 K. The non-linear region in $n_{Hall}$–$V_{GS}$ of the IZO Hall bar at 2 K in (f) is caused by free carriers and localized (trapped) carriers with different mobilities.

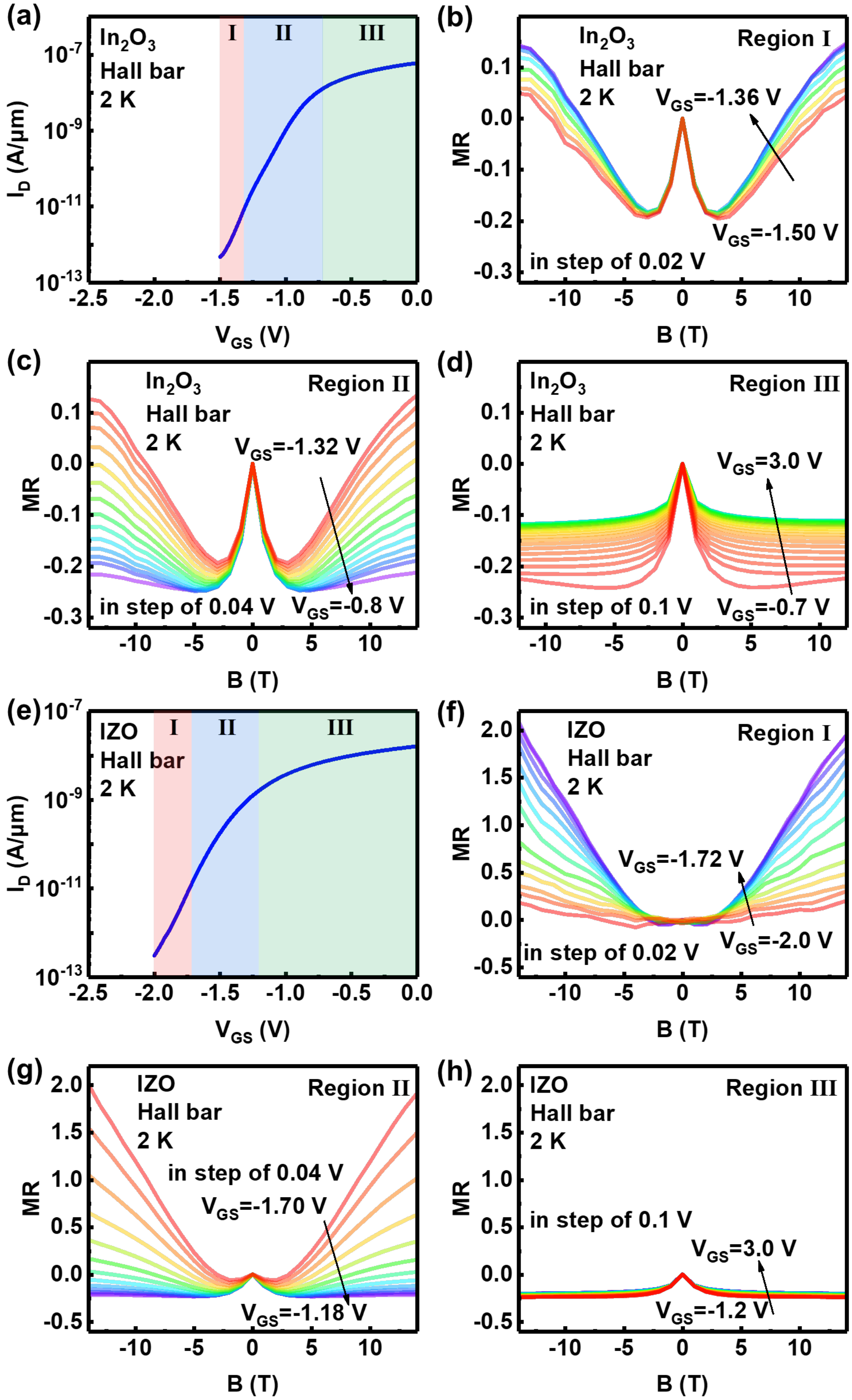

(a)
$In_2O_3$
Hall bar
2 K
I
II
III
$I_D$ (A/μm)
$V_{GS}$ (V)
(b)
$In_2O_3$
Hall bar
2 K
Region I
$V_{GS}$=-1.36 V
$V_{GS}$=-1.50 V
in step of 0.02 V
MR
B (T)
(c)
$In_2O_3$
Hall bar
2 K
Region II
$V_{GS}$=-1.32 V
in step of 0.04 V
$V_{GS}$=-0.8 V
MR
B (T)
(d)
$In_2O_3$
Hall bar
2 K
Region III
$V_{GS}$=3.0 V
in step of 0.1 V
$V_{GS}$=-0.7 V
MR
B (T)
(e)
IZO
Hall bar
2 K
I
II
III
$I_D$ (A/μm)
$V_{GS}$ (V)
(f)
IZO
Hall bar
2 K
Region I
$V_{GS}$=-1.72 V
in step of 0.02 V
$V_{GS}$=-2.0 V
MR
B (T)
(g)
IZO
Hall bar
2 K
Region II
in step of 0.04 V
$V_{GS}$=-1.70 V
$V_{GS}$=-1.18 V
MR
B (T)
(h)
IZO
Hall bar
2 K
Region III
in step of 0.1 V
$V_{GS}$=3.0 V
$V_{GS}$=-1.2 V
MR
B (T)

**Figure 4.** Transfer curves of (a) the $In_2O_3$ and (e) the IZO Hall bar at 2 K and B of 0 T, divided into three regions. MR (MR = [R(B)-R(0)]/R(0)) of the $In_2O_3$ Hall bar in (b) region I, (c) region II, and (d) region III, and of the IZO Hall bar in (f) region I, (g) region II, and (h) region III at 2 K. The positive MR occurs at high B in regions I and II for both devices, and is more pronounced for the IZO device.

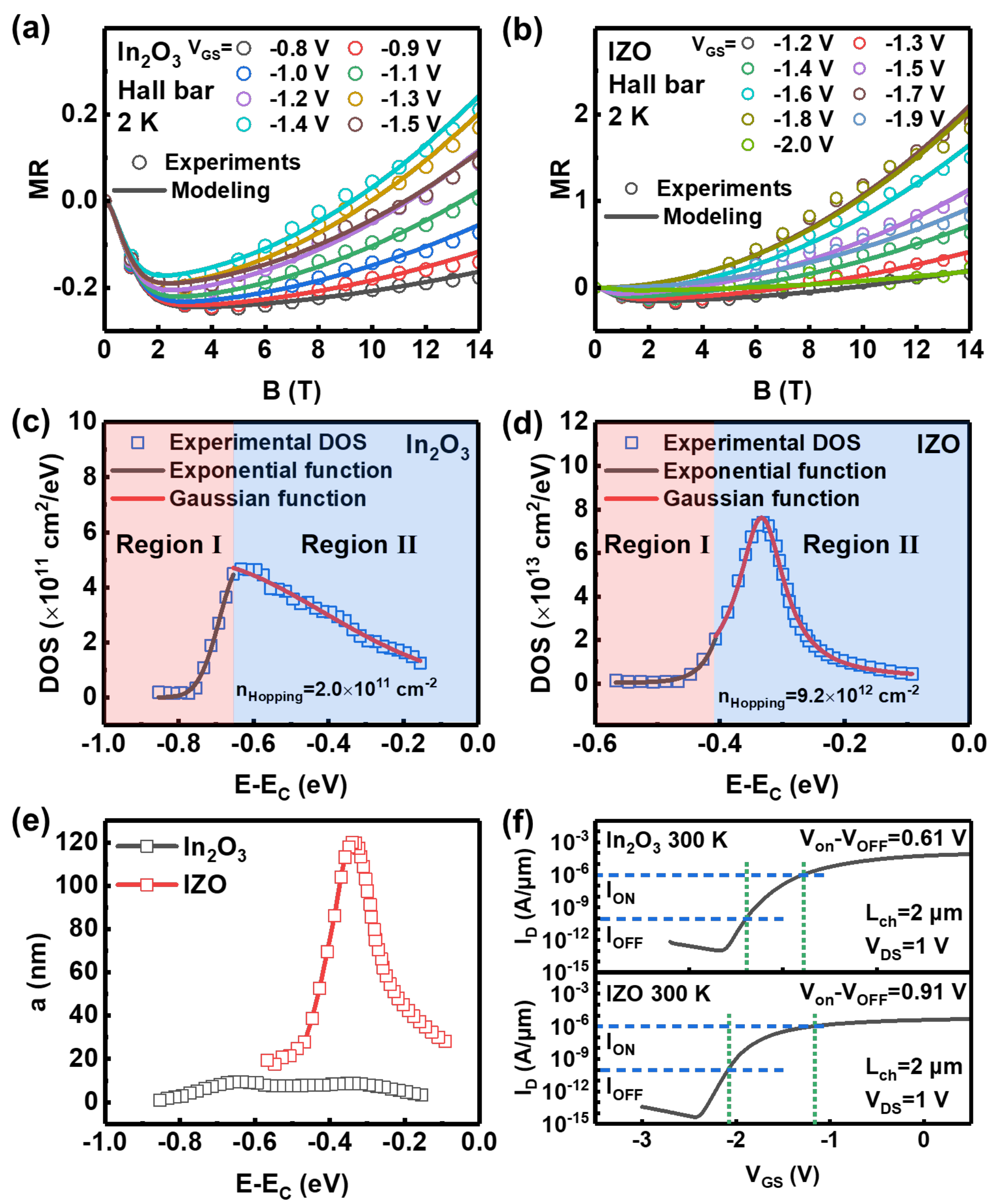

**Figure 5.** Numerical fitting of the MR of the (a) $In_2O_3$ Hall bar at $V_{GS}$ of −0.8 to −1.5 V and (b) IZO Hall bar at $V_{GS}$ of −1.2 to −2.0 V at 2 K. Fitting results of the DOS of (c) $In_2O_3$ and (d) IZO extracted from MR at 2 K. (e) Localization radius a of $In_2O_3$ and IZO extracted from MR at 2 K. (f) Comparison of the transition-region width $V_{TR}$ of the $In_2O_3$ and IZO transistors. $V_{OFF}$ and $V_{ON}$ are extracted when $I_D$ is equal to $10^{-10}$ A/μm and $10^{-6}$ A/μm, respectively.

## Acknowledgements

This work was supported by National Key R&D Program of China under Grant 2022YFB3606900, the National Natural Science Foundation of China under Grant 62274107, 92264204 and 62604209, and Shanghai Pilot Program for Basic Research-Shanghai Jiao Tong University under Grant 21TQ1400212.

## Author Contributions

C.W. conceived the idea for investigation of hopping conduction in oxide semiconductor transistors by magneto-transport measurements. C.W., L.Z., K.J., J.Z., Z.Z., X.L., and M.S. conducted all the data analysis. C.W. and M.S. co-wrote the manuscript and all authors commented on it.

## Competing interests

The authors declare no competing interests.

## Additional information

The supporting information includes: (1) the gated Hall effect measurement setup, including the measurement conditions; (2) the $V_{XX}$ and $V_{XY}$ versus $V_{GS}$ characteristics of the IZO Hall bar, from which the critical $V_{GS}$ for reliable Hall measurements is identified; (3) the two-carrier fit of the non-linear $n_{Hall}$–$V_{GS}$, including the subgap DOS model that combines an exponential band tail with a Gaussian distribution and the resulting DOS; (4) the magnetoresistance measurement configuration; and (5) the conversion of $V_{GS}$ to the surface potential $\psi_s$ through the charge balance of the gate stack, together with the DOS versus $V_{GS}$ of the $In_2O_3$ and IZO devices.

**Supporting Information**

# Investigation of Hopping Conduction and Its Impact on the Subthreshold and Transition Region Transport in Oxide Semiconductor Transistors by Magneto-Transport Measurements

Chen Wang[1], Liankai Zheng[1], Kai Jiang[1], Jinxiu Zhao[1], Zhenyu Zhang[2], Xuefei Li[2], and Mengwei Si[1,*]

[1] National Key Laboratory of Advanced Micro and Nano Manufacture Technology and School of Information Science and Electronic Engineering, Shanghai Jiao Tong University, Shanghai 200240, China

[2] School of Integrated Circuits, Huazhong University of Science and Technology, Wuhan 430074, China

* Corresponding author. Email: mengwei.si@sjtu.edu.cn

## 1. Gated Hall effect measurement setup

In the gated Hall measurement, the voltage parallel to $I_D$ ($V_{XX}$) and the Hall voltage ($V_{XY}$) were measured by OE1022 and OE1022D lock-in amplifiers, and the gate voltage ($V_{GS}$) was applied by a Keithley 2400 source-meter. An SR570 current preamplifier was used to convert the drain current ($I_D$) into a voltage, so that the $I_D$ could be measured simultaneously with the Hall voltages. The gated Hall measurement was carried out in a physical property measurement system (PPMS, DynaCool-14T). The driven AC signal has a frequency of 17.7 Hz and an amplitude of 0.1 $V_{RMS}$. [2,1]

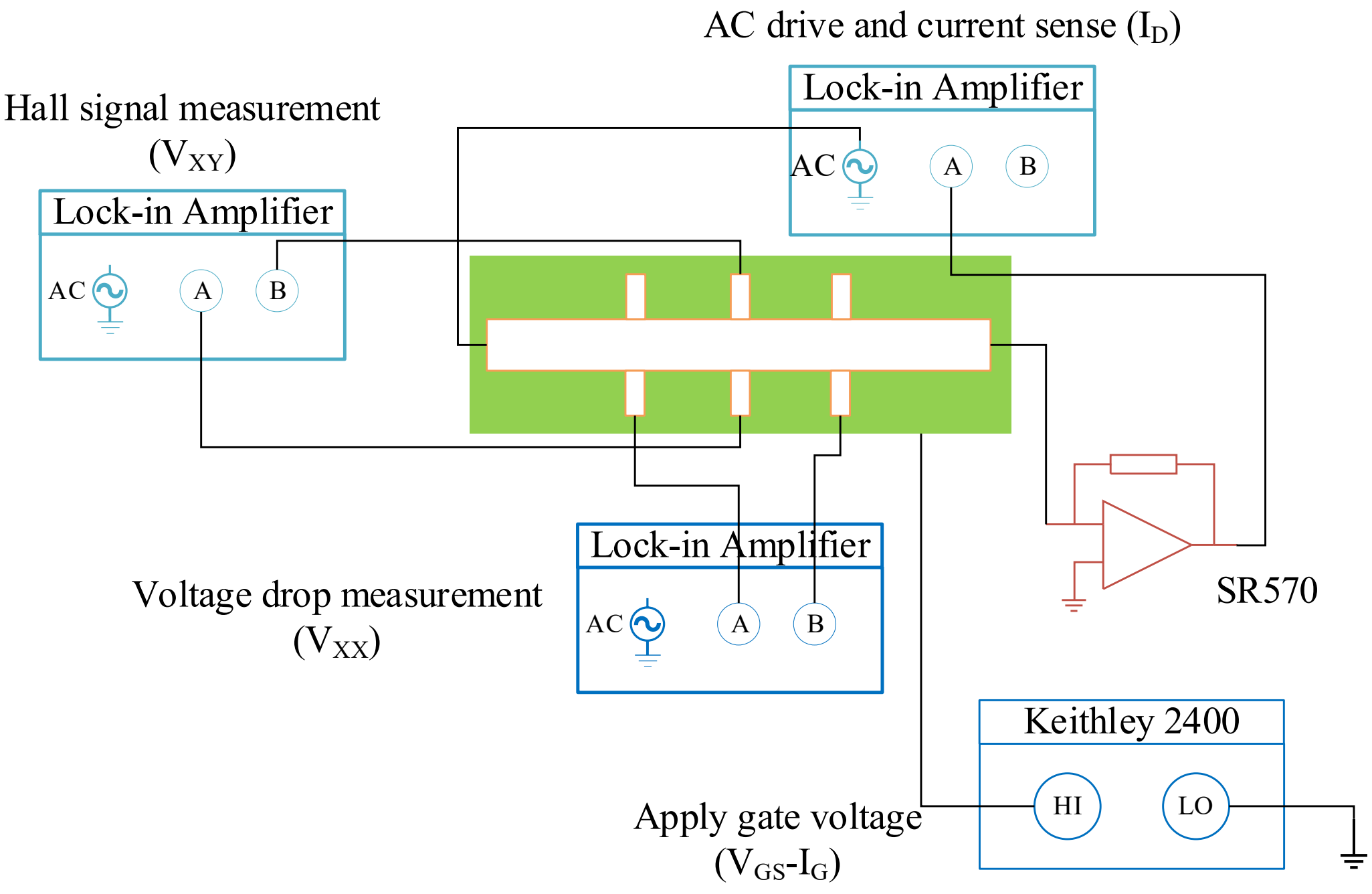

**Figure S1.** Gated Hall effect measurement setup.

## 2. Gated Hall effect measurement of IZO device

Figure S2 shows the $V_{XX}$ and $V_{XY}$ of the IZO Hall bar as functions of $V_{GS}$ at 2 K.

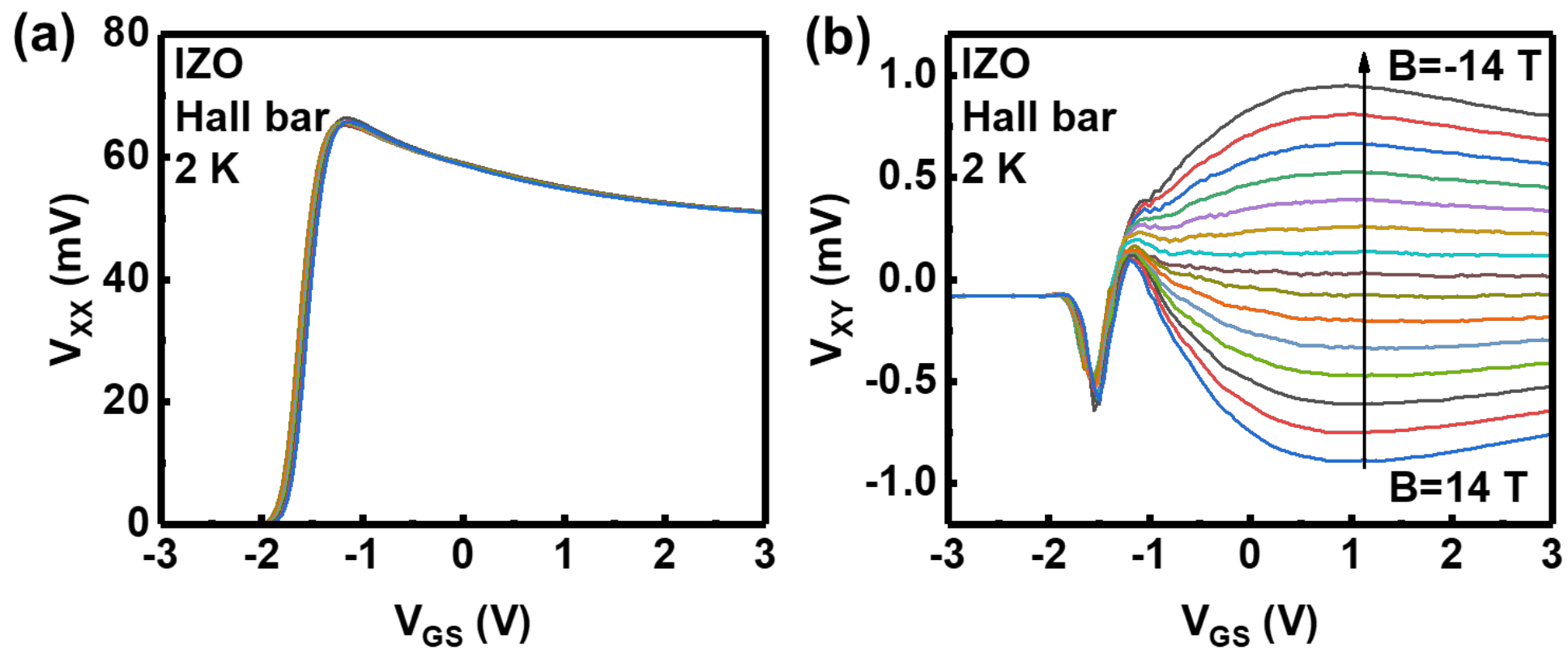

**Figure S2.** (a) $V_{XX}$ and (b) $V_{XY}$ versus $V_{GS}$ characteristics of an IZO Hall bar measured at 2 K with $B$ from −14 T to 14 T.

**3. Two-carrier fit of the Hall carrier density**

The non-linear $n_{\text{Hall}}$–$V_{GS}$ of the IZO Hall bar at 2 K (Fig. 3(f) of the manuscript) was reproduced with the two-carrier expression of eqn (1) of the manuscript, in which the induced carriers are separated into free (extended-state) electrons and localized carriers occupying subgap states. The ratio of the two mobilities enters eqn (1) of the manuscript, so $\mu_{\text{Loc}}/\mu_{\text{Free}}$ was used as the fitting parameter. As discussed in the main text, this fixed ratio is only an approximation, because $\mu_{\text{Loc}}$ and $\mu_{\text{Free}}$ are themselves energy-dependent; the DOS obtained from this fit therefore provides an approximate description of the shallow states. The total induced charge is related to $V_{GS}$ by[3]

$$V_{\text{GS}} = \frac{q(n_{\text{free}}+n_{\text{loc}}-n_0)}{C_{\text{ox}}} + \psi_s \quad \text{(S1)}$$

where $n_0$ is the equilibrium sheet carrier density, $C_{ox}$ the gate capacitance per unit area, and $\psi_s$ the surface potential, that is, the band bending measured with respect to $E_C$.

The subgap density of states was described by a Gaussian distribution of localized states joined to an exponential band tail:

$$D(E) = D_{\text{bd}}\exp\left[\frac{E-E_{\text{bd}}}{W_t}\right]\ (E < E_{\text{bd}}) \quad \text{(S2)}$$

$$D(E) = (g_L - g_{min})\exp\left[-\frac{(E-E_0)^2}{W_G^2}\right] + g_{min}\ (E \geq E_{\text{bd}}) \quad \text{(S3)}$$

where $E_0$ is the energy of the maximum, $g_L$ the peak density of states, $g_{min}$ a constant background, $W_G$ the width of the Gaussian, and $W_t$ the decay width of the exponential tail. $D_{bd}$ is chosen so that $D(E)$ is continuous at $E_{bd}$.

The two carrier densities follow from

$$n_{\text{loc}} = \int D(E)f(E)dE \quad \text{(S4)}$$

$$n_{\text{free}} = \int N_c^{2D} f(E)dE \quad \text{(S5)}$$

where $f(E)$ is the Fermi–Dirac function, $E_F$ the Fermi level, and $N_c^{2D}$ the two-dimensional conduction-band density of states. Since the thermal energy at 2 K is much smaller than the energy scale over which $D(E)$ varies, these integrals were evaluated analytically.

Both carrier densities depend on the surface potential, that is, on the Fermi level $E_F = q\psi_s$: $n_{\text{Loc}}$ through eqn (S4) and $n_{\text{Free}}$ through eqn (S5). The link between $\psi_s$ and $V_{GS}$ is provided by eqn (S1). Solving eqn (S1) for $\psi_s$ at each $V_{GS}$ therefore establishes the relation between $V_{GS}$ and $E_F$, and substituting this relation back into eqns (S4) and (S5) yields $n_{\text{Free}}(V_{GS})$ and $n_{\text{Loc}}(V_{GS})$ directly. As $V_{GS}$ increases, $E_F$ moves toward the conduction-band edge, so that the localized states are progressively filled and $n_{\text{Loc}}$ rises steeply. The crossover from a hopping-dominated to a free-electron-dominated regime obtained in this way is the origin of the non-linear $n_{\text{Hall}}$–$V_{GS}$ described by eqn (1) of the manuscript.

The parameters of $D(E)$ and the mobility ratio were then determined by a least-squares fit to the measured $n_{\mathrm{Hall}}$–V<sub>GS</sub>. The calculated curve reproduces the non-linear $n_{\mathrm{Hall}}$–$V_{GS}$ of the IZO device. The resulting $D(E)$ and the corresponding fit are shown in Fig. S3.

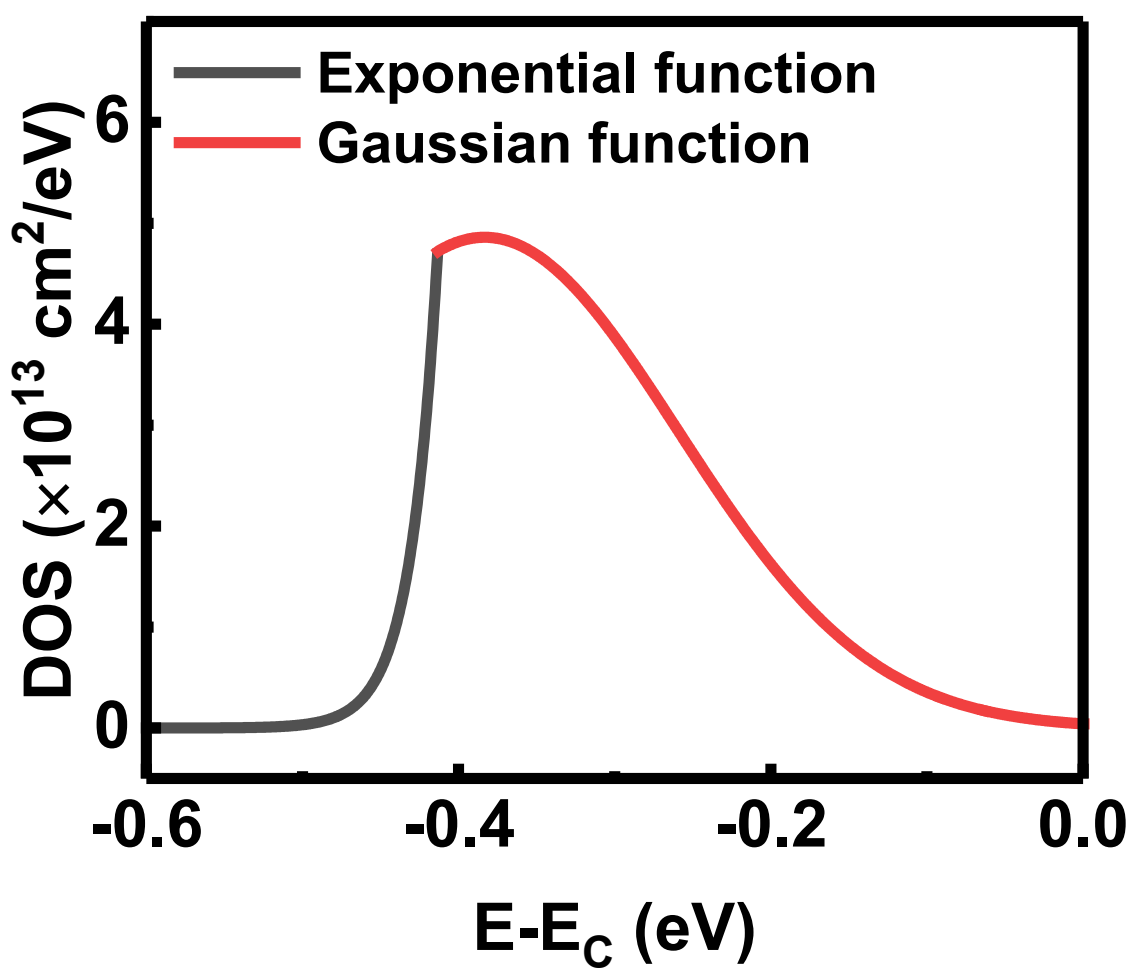

**Figure S3.** Subgap density of states $D(E)$ of the amorphous IZO channel that was used to reproduce the non-linear $n_{\mathrm{Hall}}$–$V_{GS}$ of the IZO Hall bar at 2 K.

## 4. Magnetoresistance measurement setup

Figure S4 shows the measurement configuration used for the MR measurements, which differs from that of the gated Hall measurement.

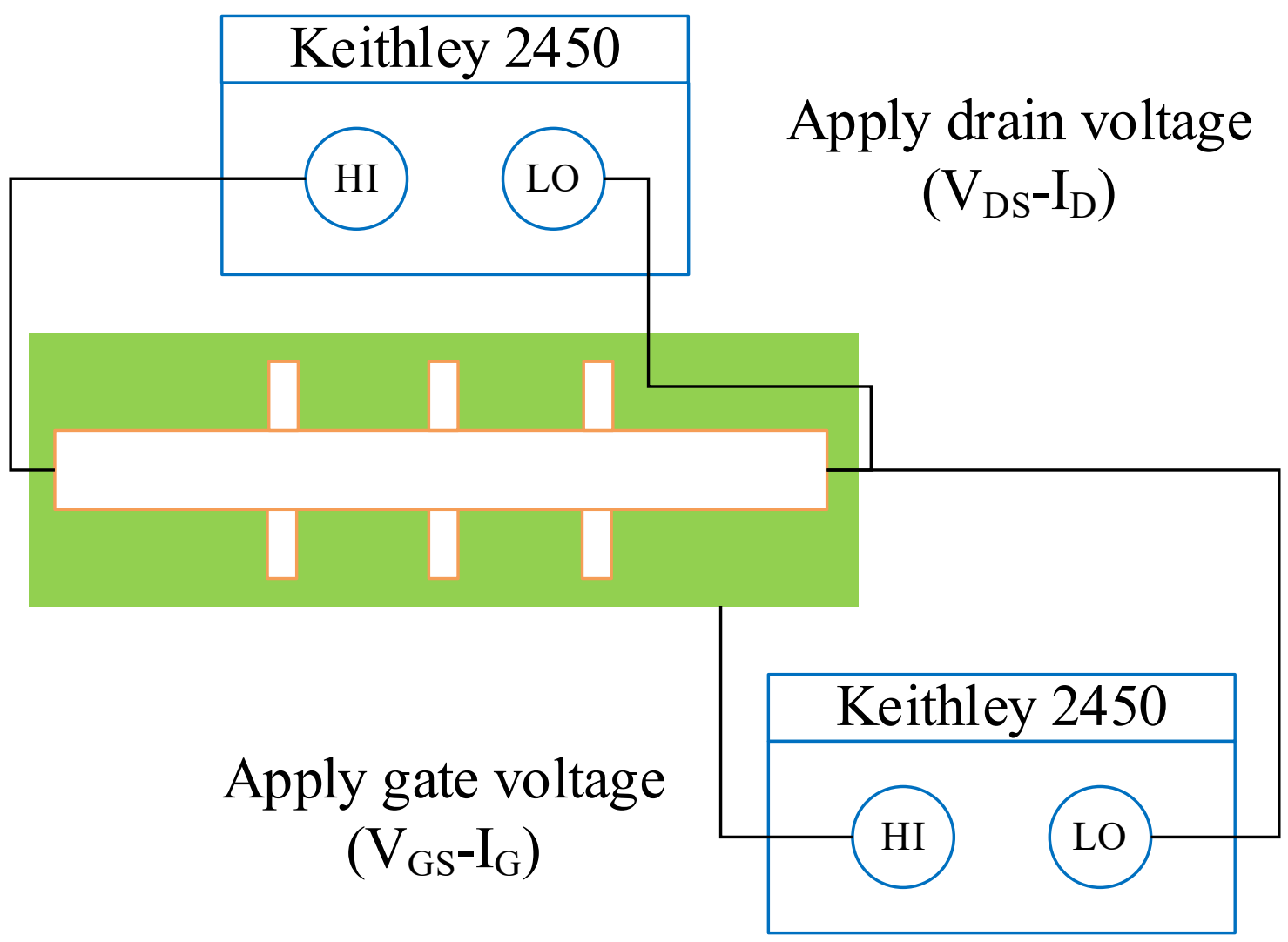

**Figure S4.** Magnetoresistance (MR) measurement on the Hall bar devices.

### 5. Mapping of the $V_{GS}$ to the surface potential

The subgap DOS obtained from the MR analysis (Fig. S5) is a function of the $V_{GS}$. To place it on an energy scale, the applied $V_{GS}$ is converted into the surface potential $\psi_s$, that is, the band bending with respect to the conduction-band edge, through the charge balance of the gate stack[3]:

$$V_{\mathrm{GS}} = \psi_s + \frac{q}{C_{\mathrm{ox}}}\int_{-\infty}^{+\infty} g(E)f(E,\psi_s)dE - \frac{q}{C_{\mathrm{ox}}}\int_{-\infty}^{+\infty} g(E)f(E,\psi_s = 0)dE \qquad \text{(S6)}$$

where $C_{ox}$ the gate capacitance per unit area, DOS the density of states, and $f(E,\psi_s)$ the Fermi–Dirac function. The second integral subtracts the equilibrium charge at zero band bending. For a two-dimensional conduction band with a constant density of states, eqn (S6) reduces to

$$V_{\mathrm{GS}} = \psi_s + \frac{q(N_c^{2D}\psi_s - n_0)}{C_{\mathrm{ox}}} \qquad \text{(S7)}$$

where $N_c^{2D}$ is the two-dimensional conduction-band density of states and $n_0$ the equilibrium sheet carrier density. For each value of $V_{GS}$, eqn (S6) is solved for $\psi_s$, and the resulting Fermi level $E_F = q\psi_s$ then gives the energy position of the states being probed, so that the DOS and the localization radius $a$ are obtained as functions of the energy $E - E_C$, as shown in Figs. 5(c) and 5(d) of the manuscript. The same charge-balance relation underlies the two-carrier fit of $n_{\text{Hall}}$ (eqn (S1)), in which the induced charge is expressed through $n_{\text{Free}}$ and $n_{\text{Loc}}$.

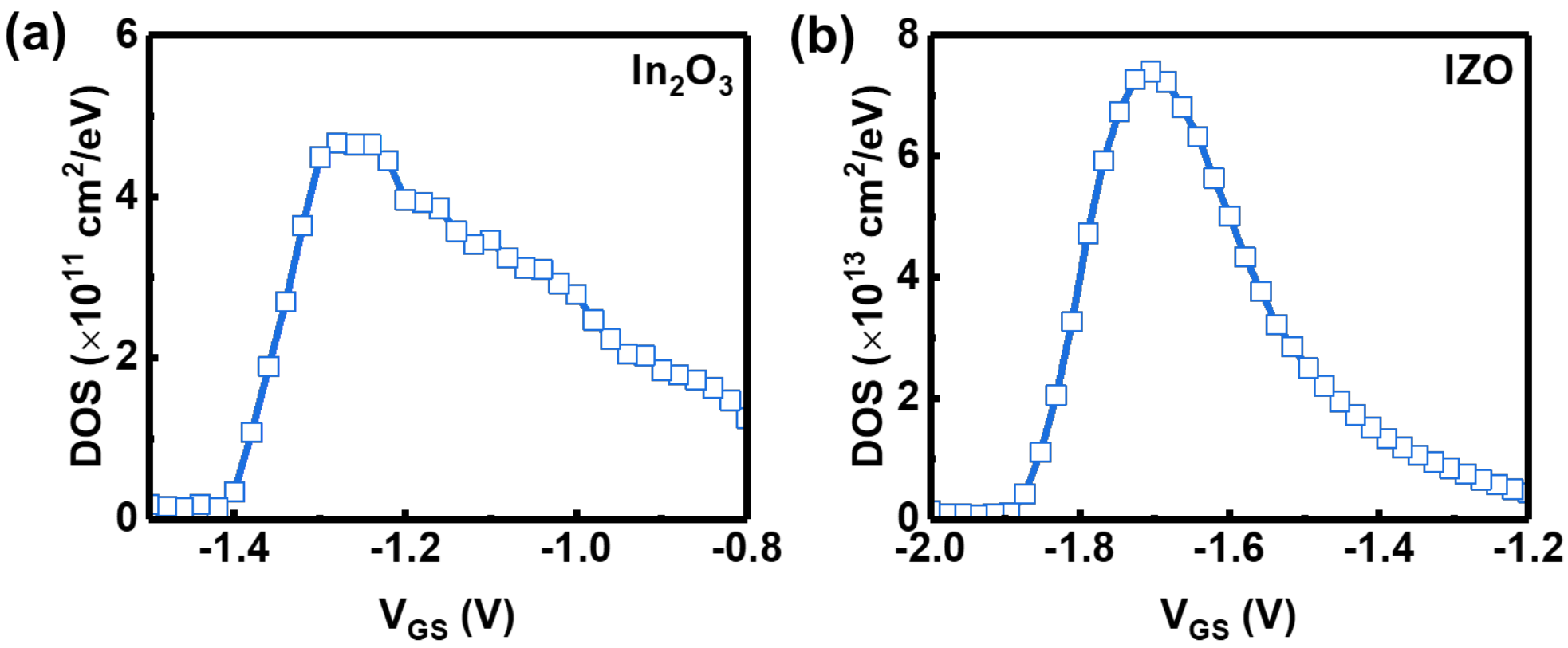

**Figure S5.** DOS versus $V_{GS}$ of (a) $In_2O_3$ and (b) IZO from MR at 2 K.